\documentclass{kapproc}
\usepackage{psfig}
\usepackage{latexsym}

\begin{document}

\articletitle{Nebular diagnostics for young\\
stellar populations: Photon\\ 
escape {\em vs} atmosphere blanketing}

\author{{\'A}ngeles I. D\'{\i}az and Marcelo Castellanos}
\affil{Departamento de F\'{\i}sica Te{\'o}rica\\
Universidad Aut{\'o}moma de Madrid, Spain}

\begin{abstract}
Emission lines in ionized nebulae can provide strong and useful
constraints on the properties of both ionizing and non-ionizing stellar
populations in regions with star formation, provided that stellar
evolution and stellar atmosphere models can be used in a selfconsistent
way. Recently, the application of these techniques has shown important
discrepancies between predicted and observed nebular spectra that point
to stellar atmosphere models of WR stars which are too energetic and/or
to a significant leakage of high energy photons. In this contribution
these two alternatives are analyzed in detail.
\end{abstract}

\section{Introduction.}
Young stars produce a great impact on their surrounding gas which gets
ionized exhibiting a characteristic emission line spectrum (ELS). The
analysis of this spectrum can be used to extract relevant information
about the stellar population ionizing the region.

Three are the parameters that characterize a stellar population: age,
metallicity and initial mass function (IMF). The metallicity can be
derived from the ELS provided the  determination of the electron temperature of
the gas can be made and the age can be constrained if spectral features
corresponding to stars in a specific evolutionary stage are
observed. Then, if a given IMF is assumed, the spectral energy
distribution (SED) of the ionizing population can be synthesized. This
spectrum, when used as input to a photo-ionization model should
reproduce the observed ELS. But, is this the case?

\section{Application of the method}

We have applied this method to three giant extragalactic HII regions
(GEHR): H13  in
NGC~628, CDT3 in NGC~1232 and 74C in NGC~4258. For these regions the
elemental abundances have been derived using ion-weighted temperatures
determined from the corresponding auroral to nebular line intensity
ratios. Details of the observations and abundance determinations can be
found in D\'\i az {\em et al.} (2000) and Castellanos, D\'\i az \& Terlevich
(2002a). All three regions show WR features in their spectra which
provides a reliable age constraint for the ionizing stellar population
from WR population synthesis models (Schaerer \& Vacca 1998). These
models consistently fit both the intensities and the equivalent widths
(EW) of the WR features under the asumptions of a single star burst and
a Salpeter IMF. 

With the derived metallicity and age for a given HII
region, we have synthesized the SED of the corresponding ionizing
clusters using the STARBURST99 code (Leitherer {\em et al.} 1999) and
then we have used the photo-ionization code CLOUDY (Ferland 1999) to
calculate the ELS. Constant density and ionization bounding have been 
assumed. The details of the procedure can be found in
Castellanos, D\'\i az \& Tenorio-Tagle (2002), whose results can be
summarized as follows: (1) WR feature intensities and equivalent widths
imply young ages for the three observed regions: 4-4.1 Myr for H13,
3.1-3.5 Myr for CDT3 and 4.5 Myr for 74C. (2) The synthesized SED
corresponding to these clusters DO NOT reproduce the observed ELS; they
result too hard. This can be seen in Figure 1 (left panel) where three SED for
CDT3 in NGC~1232 are shown. The cluster 3.3 Myr old
reproduces the observed WR features but results too hard to reproduce
the ELS. The cluster with 2.8 Myr is able to reproduce the ELS but do
not have WR stars. It is also possible to reproduce the ELS with a
combination of clusters with different ages: 2.8 Myr and 4.8 Myr. In
this case however, the intesity and EW of the WR features are
underpredicted by factors of about 15 and 25 respectively (Castellanos,
D\'\i az \& Terlevich 2002b).

\begin{figure}
\sidebyside
{\psfig{file=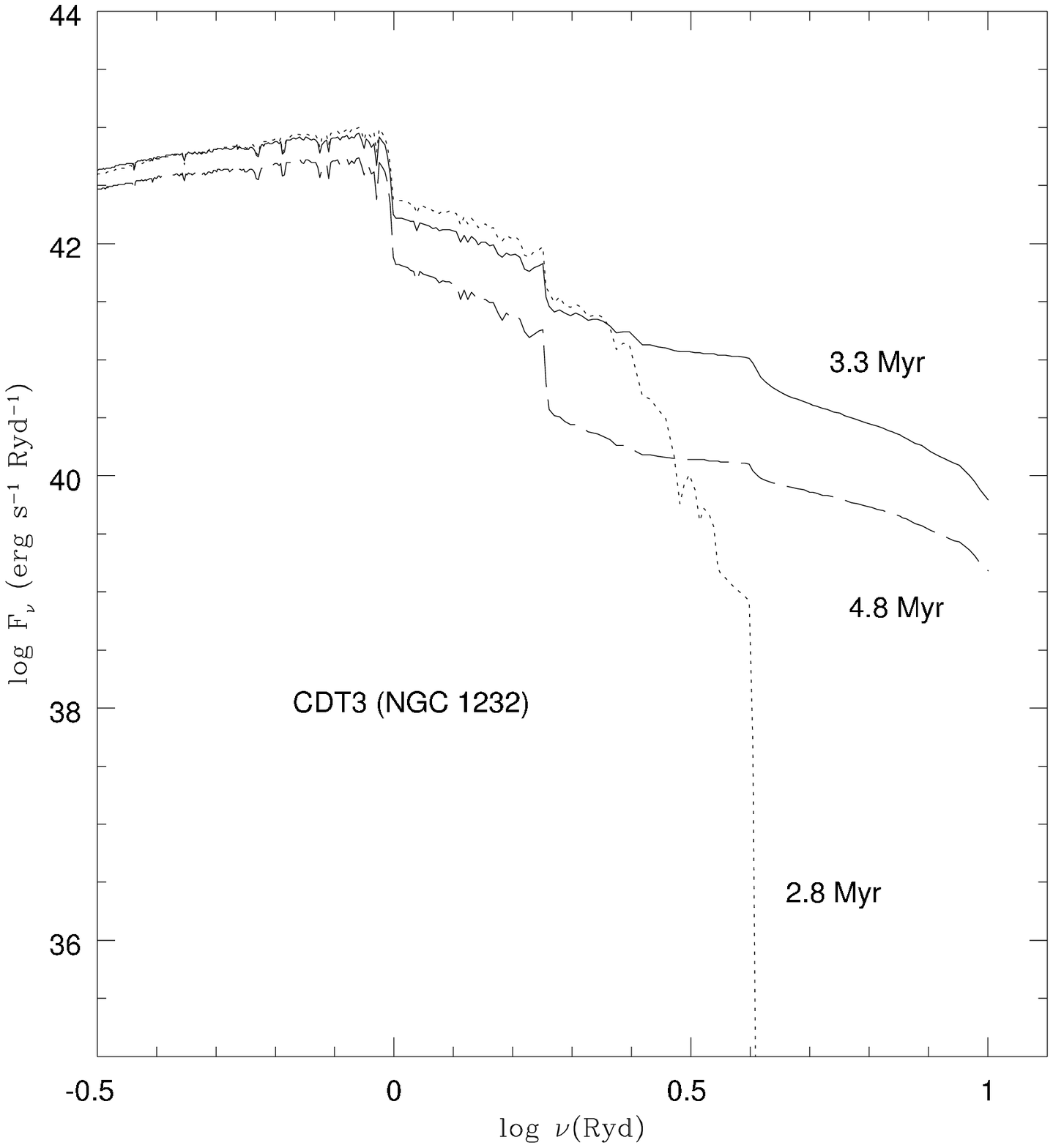,width=5cm,clip=}}
{\psfig{file=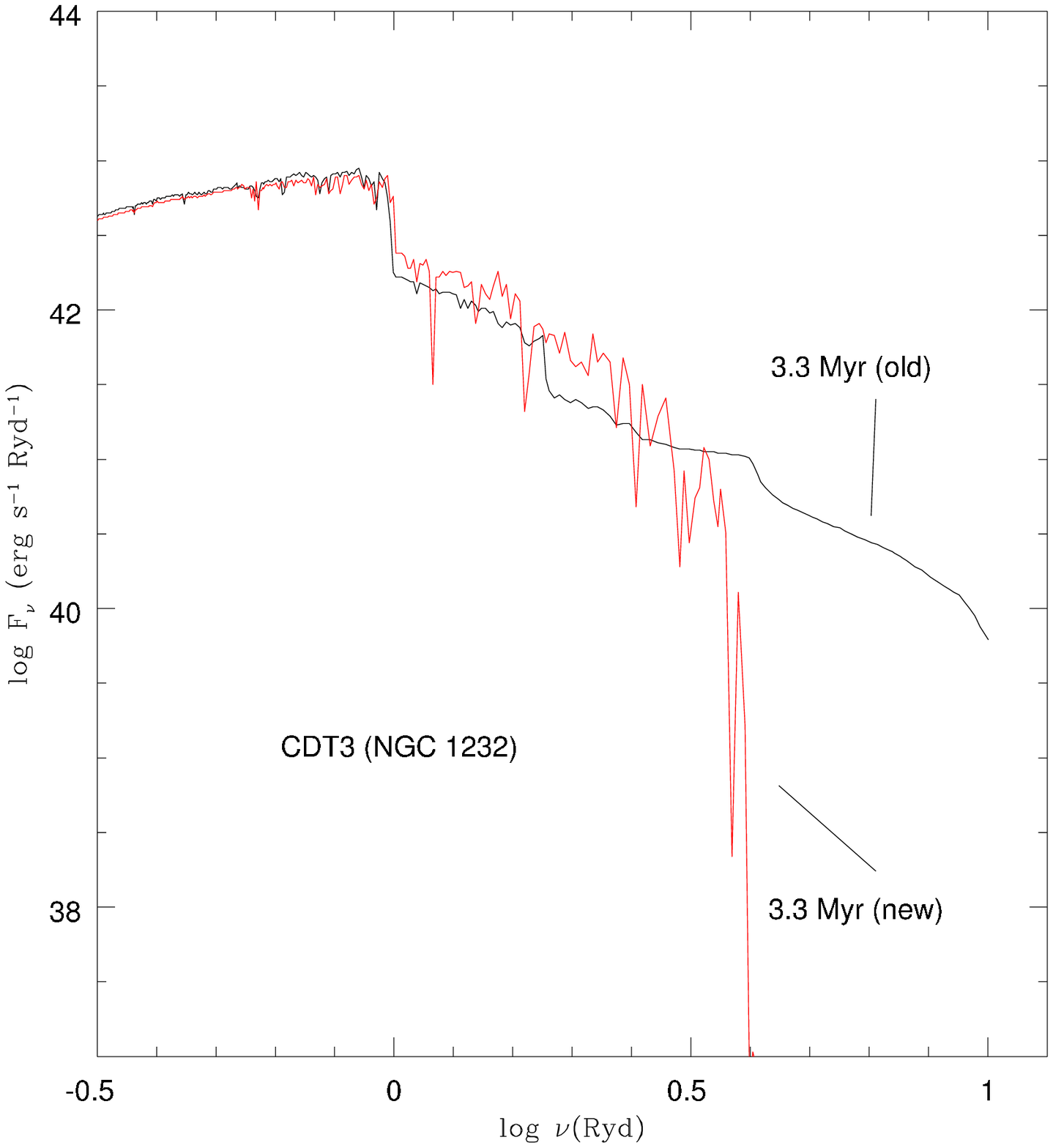,width=5cm,clip=}}
\caption{Left: SED of ionizing clusters used as input to CLOUDY to try
  to reproduce the observations of region CDT3. Right: SED of a cluster
  3.3 Myr old computed with unblanketed (old) and blanketed (new) model atmospheres}
\end{figure}

\section{Discussion}
Which could be the agents producing this discrepancy? Several
possibilities can be considered: (1) The stellar evolutionary tracks
used in the synthesis models are not correct. (2) The IMF is not the
Salpeter one. (3) The stellar atmospheres are not adequate. (4)The HII
regions are not ionization bounded.

The fact that the intensities and EW of WR features are very well
reproduced by Schaerer and Vacca's models seems to indicate that both
stellar evolutionary tracks and the assumed IMF are rather
adequate. Since these are the same assumed also by the STARBURST99
models, they cannot be the source of the discrepancy. The third
possibility is related to the uncertainties involved in the modelling of
the atmospheres of O and WR stars. Our results point to an overestimate
of the hardness of the SED of the stars in the WR phase which is more
evident for intermediate to high metallicity. The fact  that atmosphere
models for WR stars used by the STARBURST99 code do not include
blanketing seems to indicate that their assumed SED could be indeed too
hard, at least for metallicities higher than about 1/2 solar. Recently
(Smith {\em et al.} 2002) have computed models for O and WR stars
including the effects of blanketing. These models have already been
implemented in the STARBURST99 code. We have therefore synthesized the
SED of the clusters able to reproduce the WR features of our observed
HII regions using these blanketed atmosphere models. A comparison of the
``old'' and ``new'' models for the cluster of 3.3 Myr reproducing the
WR features in CDT3 can be seen in Figure 1 (right panel). Blanketed models are
indeed softer at energies higher than 3 Ryd. They are, however, still
too hard at energies between 1 and 3 Ryd thus yielding a [OIII]/H$\beta$ line
ratios which is higher than observed by a factor of about 3.5.

On the other hand, there are several indirect evidences that GEHR might
be matter bounded ({\em e.g.} Beckman {\em et al.} 2000; Collins \& Rand
2001). To test this hypothesis, we have run photo-ionization models
using as ionizing source the SED of the clusters which reproduce the
observed WR features, but relaxing the assumption of ionization
bounding (see scheme in Figure 2, left panel). The results can be seen
in Figure 2 (right panel) where the run of the
emission line intensities relative to H$\beta$, as well as the H$\beta$
luminosity and EW, as a function of the ionized shell thickness is plotted. The
observed values are shown by horizontal bars. Models in which all
photons are absorbed (most right hand points in the figure) lead to
large departures from all observed values, it is therefore
possible to find a consistent fit only if the regions are matter
bounded, {\em i. e.} if there is an important escape of ionizing
radiation from the nebula. Total values of escaping hydrogen ionizing 
photons range between 10\% and 70\% of the incident values.

\begin{figure}
\sidebyside{\psfig{file=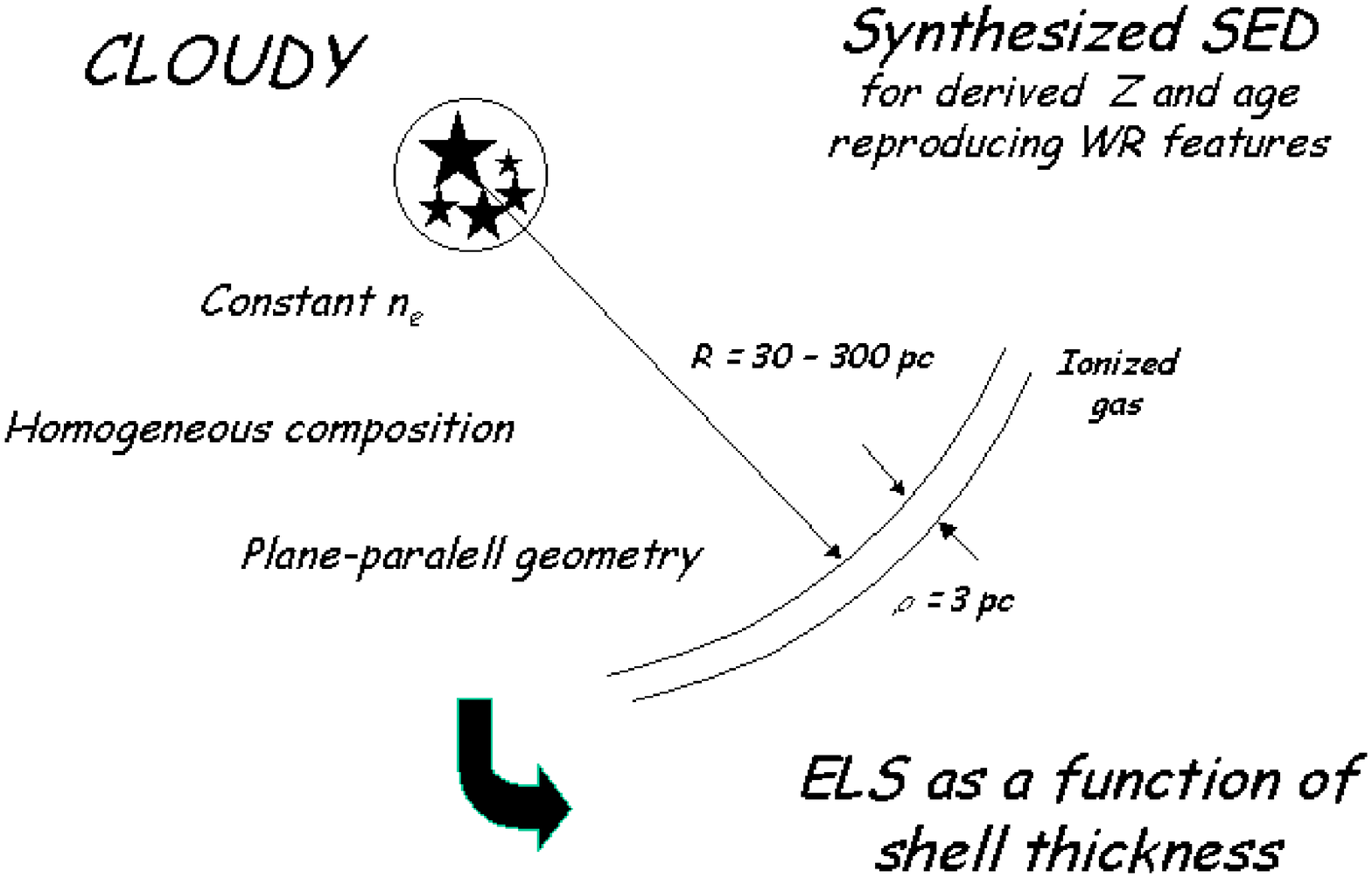,width=7cm,clip=}}
{\psfig{file=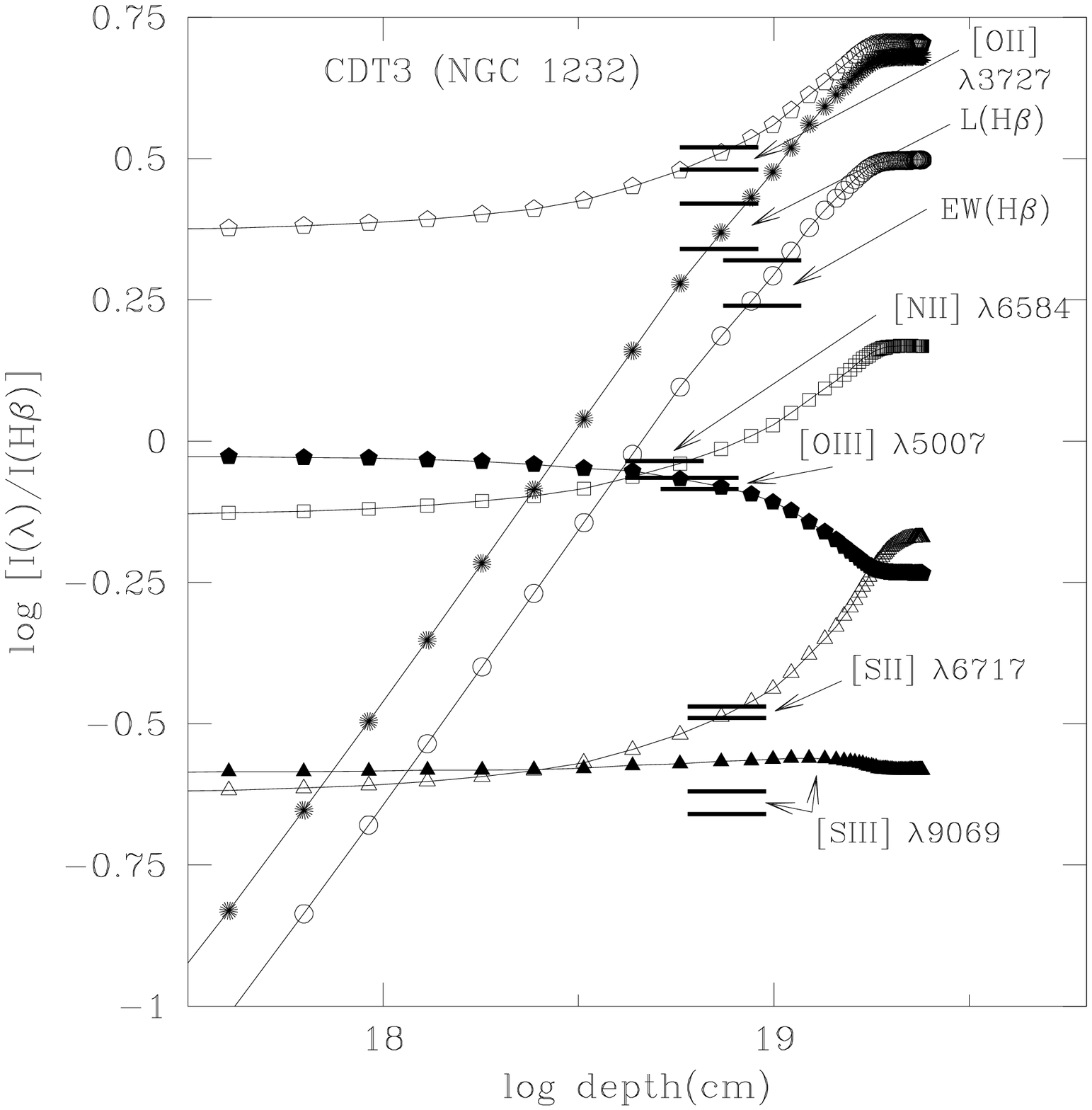,width=5cm,clip=}}

\caption{Left: Scheme of the modelling procedure.Right: Results for
  region CDT3 in NGC~1232.}
\end{figure}


This work shows (1) the potential of using nebular diagnostics to infer the
properties of the ionizing stellar populations and (2) the importance
of detecting WR features in HII region spectra in order to obtain an
independent constraint for the age of the ionizing cluster.


\begin{chapthebibliography}{}
\bibitem[]{} Beckman, J.E., Rozas, M., Zurita, A., Watson, R.A. \&
  Knapen, J.H. 2000, AJ, 119, 2728.
\bibitem[]{} Castellanos, M., D\'\i az, A.I. \ \& Tenorio-Tagle, G. 2002,
  ApJL, 565, L79.
\bibitem[]{} Castellanos, M., D\'\i az, A.I. \ \& Terlevich, E. 2002a, MNRAS,
  329, 315.
\bibitem[]{} Castellanos, M., D\'\i az, A.I. \ \& Terlevich, E. 2002b, MNRAS
  {\em in press} (Astro-ph/0208229)
\bibitem[]{} Collins, J.A. \& Rand, R.J. 2001, ApJ, 551, 57.
\bibitem[]{} D\'\i az, A.I., Castellanos, M., Terlevich, E. \ \&
  Garc\'\i a-Vargas, M.L. 2000, MNRAS, 318, 462.
\bibitem[]{} Ferland, G. 1999,HAZY: A Brief Introduction to CLOUDY,
  Univ. of Kentucky Internal Report
\bibitem[]{} Leitherer, C., Schaerer, D., Goldader, J.D. et al. 1999,
  ApJS, 123, 3. 
\bibitem[]{} Schaerer, D. \& Vacca, W.D. 1998, ApJ, 497, 618.
\bibitem[]{} Smith, L.J., Norris, R.P.F. \& Crowther, P.A. 2002, MNRAS
  {\em in press} (Astro-ph/0207554)
\end{chapthebibliography}
\end{document}